\documentclass[preprint]{revtex4}
\usepackage{graphicx}
\usepackage{bm}

\begin{document}
\title{Q-factor control of a microcantilever by mechanical sideband excitation}
\author{Warner J. Venstra} \email{w.j.venstra@tudelft.nl}
\author{Hidde J.R. Westra}
\author{Herre S.J. van der Zant}
\affiliation{Kavli Institute of Nanoscience, Delft University of Technology, Lorentzweg 1, 2628CJ Delft, The Netherlands}%
\date{\today}
\begin{abstract}
  We demonstrate the coupling between the fundamental and second flexural mode of a microcantilever. A mechanical analogue of cavity-optomechanics is then employed, where the mechanical cavity is formed by the second vibrational mode of the same cantilever, coupled to the fundamental mode via the geometric nonlinearity. By exciting the cantilever at the sum and difference frequencies between fundamental and second flexural mode, the motion of the fundamental mode of the cantilever is damped and amplified. This concept makes it possible to enhance or suppress the Q-factor over a wide range.
\end{abstract}
\maketitle
\indent\indent Cantilevers have numerous scientific and technological applications, and are used in various instruments. In sensing applications, the sensitivity is related to the Q-factor, and this has motivated researchers to increase the Q-factor of mechanical resonators, in particular in dissipative environments. Among the techniques that have been employed are applying residual stress~\cite{Craighead08}, parametric pumping~\cite{Mahboob09}, self-oscillation by internal~\cite{Steeneken11} and external~\cite{Feng08} feedback mechanisms. When increasing the Q-factor in these ways, energy is pumped into the mechanical mode and the resonator heats up. The opposite effect leads to cooling of the resonator and attenuation of its motion~\cite{Rugar91}. By pumping energy out of the mechanical resonator into a high quality-factor optical or microwave cavity, several groups have shown reduction of the effective temperature of the vibrational mode from room temperature to millikelvin  temperatures~\cite{Kippenberg08,Arcizet06,Gigan06,Schliesser09,Park09,Brown07,Regal08,Rochelau10,Poot11}. Such cooling schemes are now employed to bring down the mode temperature to below the an average phonon occupation number of one, providing a promising route to study the quantum behavior of a mechanical resonator~\cite{Cleland10, Lehnert11, Painter11}.\\
\indent\indent In analogy to cavity optomechanics, where an optical or a microwave cavity is used to extract energy from the resonator, we employ a mechanical cavity to damp the mechanical mode. Here, the fundamental flexural mode of the cantilever is the mode of interest, and the mechanical cavity is formed by the second flexural mode of the same cantilever, which is geometrically coupled to the fundamental mode. In this paper, we demonstrate the presence of this coupling by strongly driving the cantilever on resonance, while monitoring its broadband frequency spectrum. Sidebands appear in the spectrum which are located at the sum and difference frequencies of fundamental and second mode of the cantilever. Driving the cantilever at these sidebands results in positive or negative additional damping, which is demonstrated in this paper.\\
\begin{figure}[h]
\includegraphics [width=100mm] {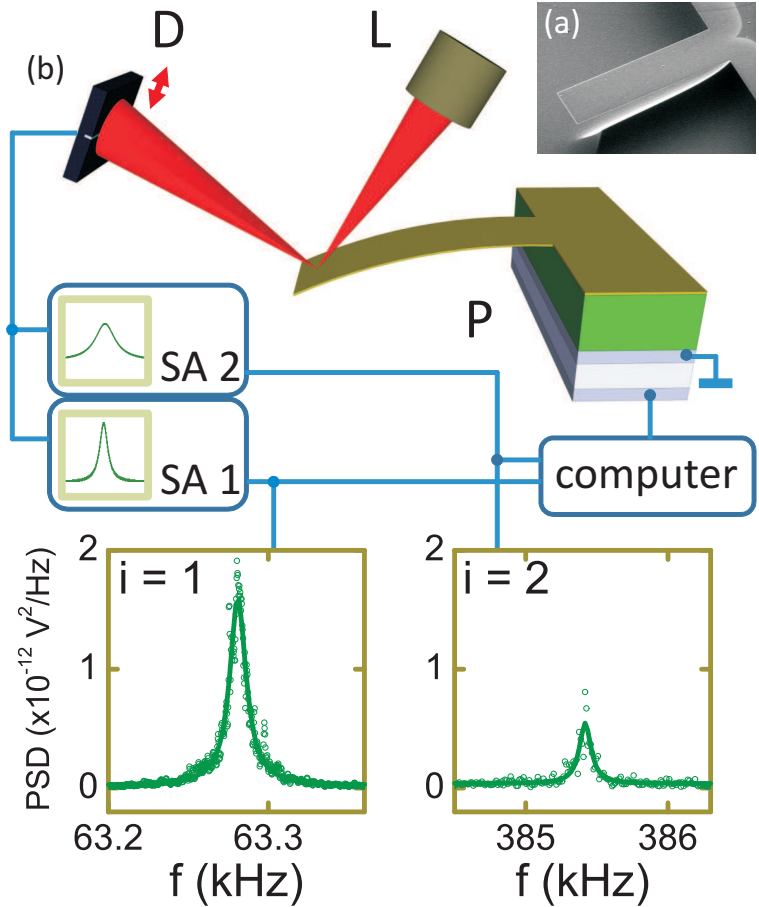}
\caption{(a) Scanning electron micrograph of the silicon nitride cantilever. (b) Diagram of the measurement circuit showing photodiode (D), laser (L), piezo (P) and the spectrum analyzers (SAs) to measure the fundamental (SA 1) and the second (SA 2) flexural mode. The thermal noise spectra are shown at the fundamental (i=1) and second (i=2) flexural mode of the cantilever.}
\end{figure}
\indent\indent Cantilevers are fabricated from low pressure chemical vapor deposited silicon nitride by electron beam lithography and isotropic reactive ion etching in a O$_2$/CHF$_3$ plasma~\cite{Gavan09JMM}. The dimensions are length $\times$ width $\times$ height $= 39\,\mathrm{\mu m} \times 8\, \mathrm{\mu m} \times 70\, \mathrm{nm}$. An optical deflection technique, similar to the one employed in atomic force microscopy, is used to detect the cantilever motion. Figure 1 (a) and (b) show the cantilever and the setup. The cantilever is mounted on a piezo crystal and placed in a vacuum chamber at a pressure of $\mathrm{\sim10^{-5}\,mbar}$. Two spectrum analyzers are used to simultaneously measure the thermal motion of the fundamental (i=1) and second (i=2) flexural mode. Figure 1(b) shows the power spectra without driving the piezo. The resonance frequencies and Q-factors are determined by fitting Lorentzian functions (solid lines), and we find  $f_1=63.2\,\mathrm{kHz}$ and $f_2=385.4\,\mathrm{kHz}$ and $f_3=1.068\,\mathrm{MHz}$ (not shown). The ratio's $f_2/f_1=6.1$ and $f_3/f_1=16.9$ are close to the expected modal frequencies $\alpha_{21}=6.3$ and $\alpha_{31}=17.5$ representing the spectrum of a homogeneous cantilevered Euler-Bernoulli beam. For the fundamental and second resonance modes, the corresponding Q-factors are $Q_1=5184$ and $Q_2 = 3922$. The frequency difference, $f_2-f_1=322\,\mathrm{kHz}$, exceeds the bandwidth of the modes, $f_1/Q_1=12\,\mathrm{Hz}$ and $f_2/Q_2=98\,\mathrm{Hz}$, by four orders of magnitude.\\
\indent\indent To demonstrate the coupling between the fundamental and second flexural modes of the cantilever, we drive the cantilever on resonance, while measuring its broadband spectrum. Figure 2(a) shows this spectrum as a function of the drive strength. When the amplitude of the second mode increases, mechanical sidebands become visible in the spectrum. These sidebands occur at $f_2\pm f_1$, and clearly indicate the presence of mechanical coupling between the two modes. Traces for weak and strong driving are extracted from (a) in Fig. 2(b), to show the shape and relative amplitudes of the sidebands. As the spacing between the sidebands is much larger than the linewidth of the mode, we operate in the resolved sideband regime~\cite{Kippenberg08}.\\
\begin{figure}[h]
\includegraphics [width=100mm] {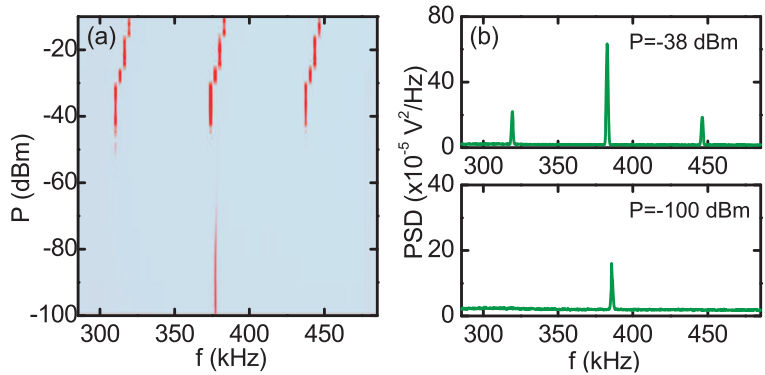}
\caption{(a) Noise spectrum while driving the second flexural at increasing amplitudes. At strong driving, sidebands emerge in the spectra at the sum and difference frequencies of the fundamental and second flexural mode. Color scale represents the power spectral density. (b) The cross-sections of panel (a) at weak (bottom) and strong (top) driving show the shape and intensity of the sidebands.}
\end{figure}\\
\indent\indent The mechanism that couples the vibrational modes in a cantilever can be qualitatively understood as follows. A nonzero amplitude of one flexural mode of the cantilever changes the shape of the cantilever~\cite{Crespo78}. This geometric change has a small but measurable effect on the resonance frequency of all the other vibrational modes. The effect of the cantilever amplitude on its own resonance frequency was recently analyzed in detail~\cite{Venstra10}: for the first few modes any nonzero amplitude stiffens the frequency response, and this gives rise to frequency pulling. Recently, we also presented a detailed study on the coupling mechanism between the vibrational modes in clamped-clamped resonators~\cite{Westra10}. Here, the coupling between the modes is fully described by the displacement-induced tension. A similar analysis can be carried out for the coupling between vibration modes of a cantilever beam. The only difference is that in the inextensional cantilever the modes are coupled by the geometric nonlinearity, whereas for the (extensional) clamped-clamped resonator the modes are coupled by the displacement-induced tension. For a cantilever, the modal amplitudes $u_i$ are calculated by solving the (dimensionless) coupled equations~\cite{Crespo90}
\begin{equation}
\ddot{u_i} + \eta_i\dot{u_i} + \omega_i^2 u_i  + \sum_{j=1}^n \sum_{k=1}^n \sum_{l=1}^n \Big(\alpha_{ijkl} u_j u_k u_l + \frac{1}{2}\beta_{ijkl} u_j (u_k u_l)\ddot{\,} \Big) = f_i \cos(\Omega_i t),\\
\end{equation}
where $\eta_i$ represents a damping constant, $\omega_i$ the resonance frequency, $\Omega_i$ the drive frequency and $f_i$ the excitation strength of mode $i$. The dots and primes denote derivative to time and coordinate, $s$, respectively. The coupling coefficients $\alpha_{ijkl}$ and $\beta_{ijkl}$ are calculated by integrating the cantilever modeshapes $\xi_i$ as follows:
\begin{eqnarray}
\alpha_{ijkl} &=& \int_0^1 \xi_i\{\xi_j'(\xi_k'\xi_l'')'\}' \mathrm{d}s \\
\beta_{ijkl} &=& \int_0^1 \xi_i \Bigg( \xi_j' \int_1^{s} \int_0^{s_2} \xi_k' \xi_l' \mathrm{d}s_1 \mathrm{d}s_2 \Bigg) ' \mathrm{d}s.
\end{eqnarray}
Taking only the fundamental and the second mode into consideration, Eq. 1 yields two coupled nonlinear differential equations with constant coefficients, which can be solved numerically.\\
\begin{figure}[h]
\includegraphics [width=100mm] {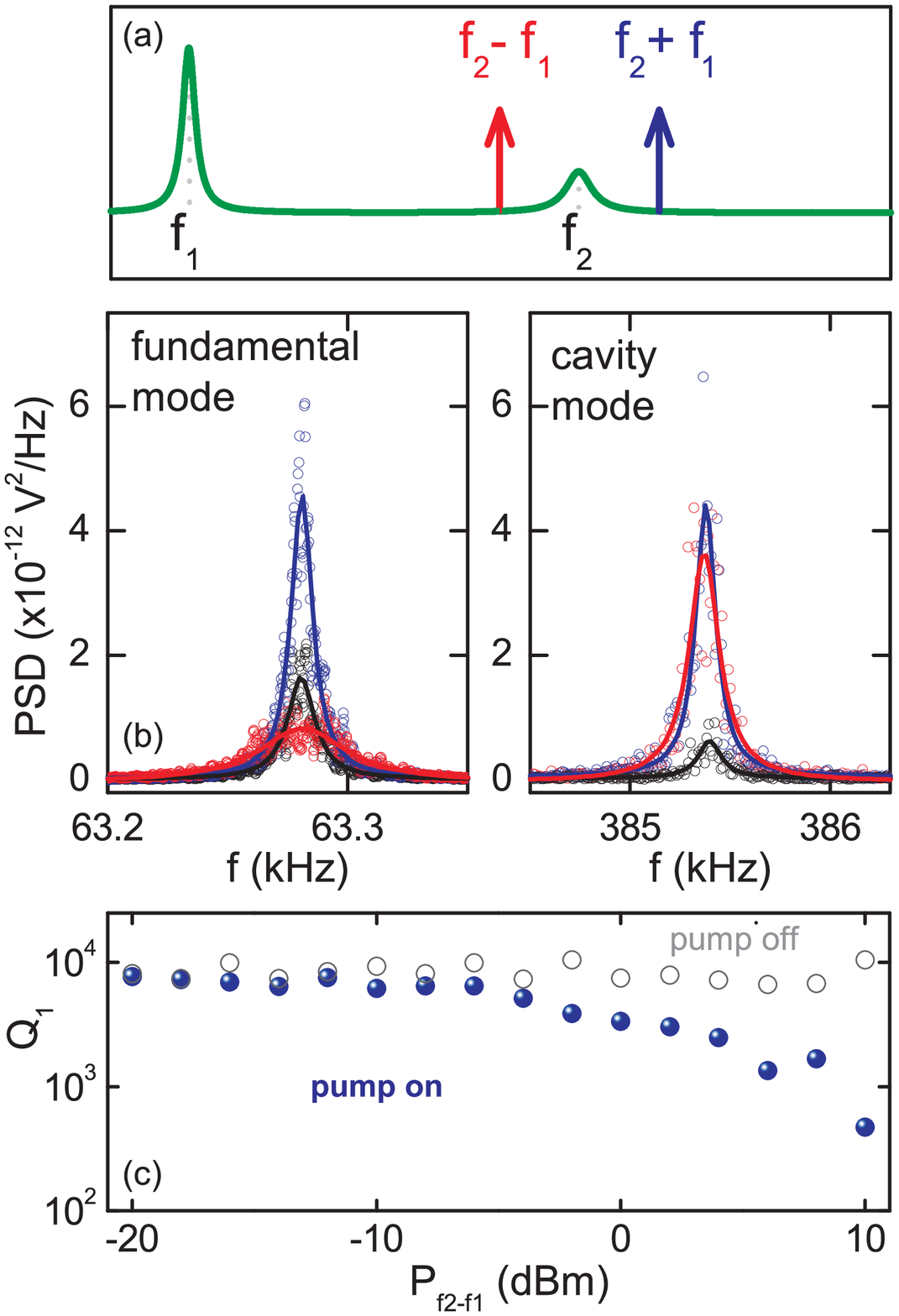}
\caption{(a) Damping and amplification of the fundamental mode by pumping the sidebands of the second flexural mode. The sum and difference frequencies are indicated by the arrows. By exciting the cantilever on the red sideband the fundamental mode is suppressed, and its motion is amplified by exciting on the blue sideband. (b) Noise spectra of the fundamental mode (left) and the cavity mode (right). The black curves represent the thermal noise spectra without excitation. The red curves are obtained by pumping the red sideband, resulting in positive damping of the cantilever. The blue curves are measured while pumping the blue sideband, which  results in negative damping (amplification). (c) The Q-factor of the fundamental mode as a function of pump power on the red sideband (closed dots). For each power a control experiment is carried out without excitation, indicated by the open circles.}
\end{figure}
\indent\indent The coupling between the vibrational modes can be used to transfer energy by employing a process similar to sideband cooling in cavity-optomechanics, where the cavity is used extract energy from the mechanical mode. The mechanical resonator is embedded in an optical~\cite{Kippenberg08,Arcizet06,Gigan06,Schliesser09,Park09} or microwave cavity~\cite{Brown07,Regal08,Rochelau10}. In analogy to those experiments and given the presence of the mechanical mode-coupling, the damping of one mechanical mode by another mode of the same resonator can be envisioned. Using the coupling mechanism described in the previous section, any change in the position of the mode under consideration (the fundamental flexural mode in the experiments that follow) changes the stiffness of the mode that acts as the cavity (the second flexural mode). The energy change in the cavity mode is retarded by the cavity relaxation time, equal to $\sim Q_2/f_2$ for our mechanical cavity. Due to the delayed response of the cavity mode, a force is exerted by the cavity mode on the fundamental mode. This velocity-proportional force can either amplify or attenuate the motion of the fundamental mode ~\cite{Schliesser08}. In case of red-detuned driving, the damping force on the fundamental mode is increased. When the driving is blue-detuned, the motion of both the cavity mode and the fundamental mode is amplified. The schemes are illustrated in Fig. 3(a), where the two Lorentzian shaped curves represent the two flexural modes of the cantilever, and the driving frequencies corresponding to blue and red detuning are indicated by the arrows. The damping rate is maximized by driving at the sum and difference frequencies, and is increased by decreasing the linewidth of the cavity mode.\\
\indent\indent The effect of sideband excitation on the damping of the cantilever is demonstrated by measuring the thermal noise spectra of the fundamental and second flexural resonance modes, while driving the piezo sinusoidally at their sum and difference frequencies. Figure 3(b) shows the spectrum without driving (indicated by the black open circles). When the cantilever is driven at the blue-detuned sideband, its amplitude increases as shown by the blue curve. The blue and red curves in the power spectral density plots of Fig. 3(b) correspond to driving at the blue and red-detuned sidebands of the cavity mode shown in Fig. 3(a). By fitting Lorentzian functions to the data, we obtain the temperature and the Q-factors of the fundamental mode while driving the sidebands. When the cantilever is driven at the red sideband, the Q-factor of the fundamental mode decreases from 4599 to 1421. No changes in the temperature of the mode are observed, which indicates that the energy extracted via the modal interactions leaks back into the mode via other transport mechanisms, which are absent in e.g. opto-mechanical cooling schemes. When driving at the blue sideband, the Q-factor increases to 5849. For the cavity mode, by red-detuned driving the Q-factor decreases from 2776 to 2108, while for blue-detuned driving it increases to 3185. Here we do observe a change in temperature, by a factor of 3.6 for the red and 6.3 for the blue-detuned driving.\\
 \indent\indent By increasing the drive strength at the red-detuned sideband the amplitude of the cantilever motion is further attenuated, as is shown in Fig. 3(c). Here, the Q-factor of the fundamental mode is shown as a function of the applied driving power at $f_2-f_1$. A 20-fold reduction of the Q-factor is achieved compared to the Q-factor without driving the sideband. This clearly demonstrates that driving at the  mechanical sidebands can be used to modify the damping characteristics of a micromechanical resonator to great extent. This scheme can be used to modify the Q-factor in cantilever-based instrumentation, where we note that the changes in damping obtained in these experiments are of the same order as the viscous damping in air, so that stronger excitation is needed to obtain a significant change in the damping.\\
\indent\indent In conclusion, we demonstrate the coupling between the flexural modes of a microcantilever. This coupling is marked by mechanical sidebands in the frequency spectrum, which are located at the sum and difference frequencies. Driving the cantilever at these mechanical sidebands results in additional damping of the resonator, which can be either negative or positive in sign. This is demonstrated for the fundamental and the second flexural mode. Furthermore, using a second mode of the same resonator as a cavity provides a means to cooling experiments based on modal interactions. In present sideband-cooling experiments, coupling a mechanical resonator to an optical or microwave cavity can pose significant experimental challenges. The coupling described in this work is present by nature, and its strength can be tuned by engineering stress and geometry. More explicitly, in  carbon nanotube resonators with extremely high Q-factors~\cite{Huettel09} at low temperatures, coupling between the vibrational modes as described in Ref.~\cite{Westra10} may provide a route to cool mechanical modes to the quantum ground state.\\
\indent\indent The authors thank Samir Etaki for discussions. Financial support from the Dutch organization FOM (Program 10, Physics for Technology) is acknowledged.

\end{document}